\documentclass[aps,prl,reprint]{revtex4-1}
\usepackage{setspace}
\usepackage{graphicx,float}
\usepackage{braket}
\usepackage{amsmath,mathtools,amssymb,mathrsfs,bm}
\usepackage{breqn}
\usepackage{adjustbox} 
\usepackage{xcolor,bm}

\begin{document}

\title{Kerr Soliton Generation in Ultra-Compact Photonic Devices}

\author{
Garrett J. Beals$^{1}$, 
Yun Zhao$^{1,3}$,
Sai Kanth Dacha$^{1}$,
James Eckstein$^{1}$,
Karl McNulty$^{2}$,
Michal Lipson$^{2}$,
Alexander L. Gaeta$^{1,2,*}$,
\\
{\small $^{1}$ Department of Applied Physics and Applied Mathematics, Columbia University, New York, New York 10027, USA}\\
{\small $^{2}$ Department of Electrical Engineering, Columbia University, New York, New York 10027, USA}\\
{\small $^{3}$ Department of Applied Physics and Ginzton Laboratory, Stanford University, Stanford, CA 94305, USA}
}
\date{\today}

\begin{abstract}
Chip-based nonlinear photonics offer the capability to integrate devices with all the requisite photonic components (e.g., filters, couplers, detectors) into highly compact form factors. 
This offers the possibility of making the devices scalable, robust, and manufacturable. 
Such integrated photonic devices will enable applications in optical communications, precision metrology, microwave generation, and LIDAR. However, thermal instabilities represent a major hurdle in the deterministic operation of nonlinear optical processes in such integrated resonant structures such as microresonators. 
In this work we demonstrate deterministic and highly stable Kerr soliton comb generation in tight-spiral microresonators with comb spacings as low as 16 GHz. 
We perform a comprehensive experimentally-validated thermal model of such compact microresonators and reveal non-trivial thermally-driven instabilities governing the cavity soliton dynamics. 
We design and implement a fast feedback loop on the devices to overcome thermal perturbations and stabilize cavity-soliton states, including those that are otherwise unstable, and to allow for controlled transitions between the soliton states. 
Our approach enables the realization of thermally-stable highly compact soliton microcomb devices in a wide variety of photonic platforms.
\end{abstract}

\maketitle

\onecolumngrid
\section{Main}
Since their discovery over ten years ago \cite{Herr2014}, dissipative Kerr solitons (DKS) in integrated microresonators have driven significant progress in nonlinear optics, offering a compact and scalable approach to comb generation from a single-frequency laser.
These combs offer a wide-range of applications including coherent optical communication \cite{Marin-Palomo2017}, LIDAR \cite{Riemensberger2020}, parallel processing \cite{Feldmann2021}, self-referencing \cite{selfref}, spectroscopy \cite{spectroscopy,spectropscopy2,suh_searching_2019,obrzud_microphotonic_2019}, and frequency synthesis \cite{Spencer2018}.
When the repetition rate is electronically detectable, integrated Kerr combs enable the generation of low-noise microwaves \cite{Kudelin2024,Sun2024,ofd}.
Such compact, low-noise microwave sources are directly relevant for satellite navigation (GPS), Doppler and synthetic-aperture radar, automotive radar, deep-space navigation, and wireless communication systems operating in the X- and Ku-bands (10-18 GHz). 
The advantage of a photonics-based approach is to reduce the size, weight, power, and cost (SWaP-C), and to increase the scalability of microwave systems.
Generating low-noise Kerr-comb-based microwaves in sub-20 GHz regime has been realized in whispering gallery mode (WGM) resonators made out of $\text{SiO}_2$ \cite{Yi:15} and $\text{MgF}_2$ \cite{Herr2014,Liang2015}. 
It is challenging to achieve such repetition rates in planar integrated devices \cite{yang_bridging_2018,Liu:18}, and has been demonstrated in fully integratable and planar platforms like silicon nitride (SiN) microresonators \cite{liu_photonic_2020,Xuan:16,ye_integrated_2022}. 
In planar-integrated microresonators, the cavity length required for sub-20 GHz microwave frequencies is on the order of 1 cm for SiN, which, for a simple ring resonator geometry corresponds to a radius of more than 1.6 mm.
Microresonators of such large radii do not take advantage of the potential compactness that planar integration provides.
For high-confinement SiN---with thicknesses $>$ 600 nm---recent work has demonstrated microwave generation in a $5 \times 5 \text{ mm}^2$ footprint \cite{liu_photonic_2020}.
However, it remains challenging to achieve low-$f_{\text{rep}}$ (repetition-rate frequency) soliton generation a 1 x 1 $\text{mm}^2$ footprint \cite{Xuan:16,ye_integrated_2022}. 

\begin{figure*}[htbp]
	\centering\includegraphics[width=\textwidth]{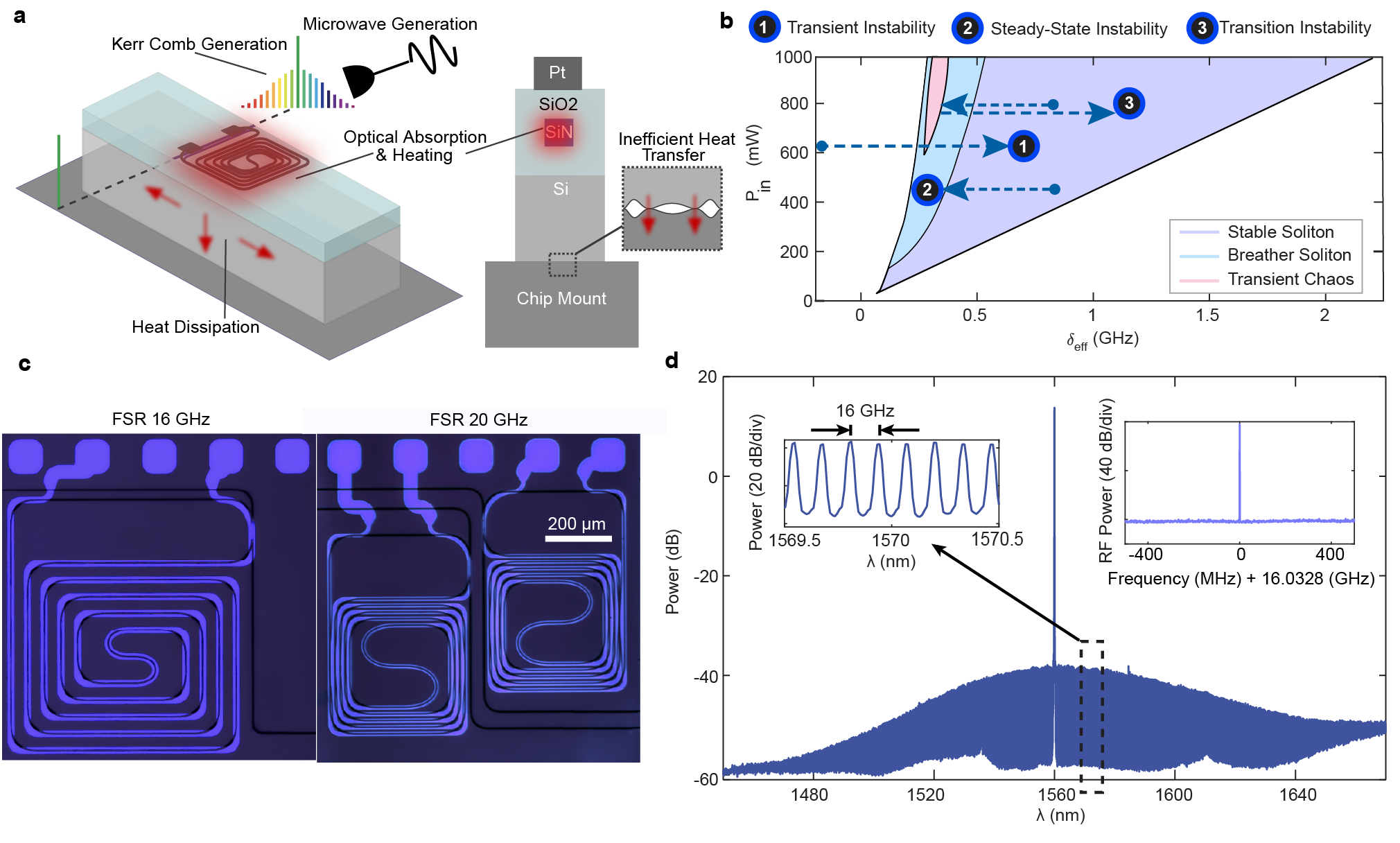}
	\caption{
		\textbf{Schematic illustrating Kerr comb generation and heat dynamics in an integrated spiral microresonator. }
		(a) A CW laser is sent into a low-FSR spiral microresonator and initiates Kerr comb generation which can be used for low-noise microwave generation. 
		Some of the optical power is absorbed and converted to heat, which dissipates across the entire chip. 
		The resulting change in temperature causes a change in the resonance detuning which strongly impacts soliton generation and stability.
		The right shows a 2D cross-sectional view of the photonic chip (not to scale). The interface between the silicon substrate and chip mount is not-ideal, and results in inefficient heat transfer. 
		(b) Soliton existence map. 
		States included are the stable soliton solutions, breather-soliton solutions, and the transient chaos regime. The last refers to the region where there is only a CW attractor and the soliton solution will decay to the CW solution. The three types of instabilities discussed in this paper are shown schematically through the dashed arrows. 
		(c) Microscope pictures of Euler spiral microresonators with a free-spectral range of approximately 16 GHz (left) and 20 GHz (right).
		(d) Single-soliton optical spectrum for a 16 GHz spiral microresonator device.
		The left inset shows the comb spacing and the right inset shows an example RF spectrum with a narrow 16 GHz microwave tone.
	}
	\label{fig:Intro}
\end{figure*}

Thermal-related instabilities in DKS generation date back to the initial demonstrations \cite{Herr2014} and, over 10 years later, remain outstanding problems \cite{mass_manu,copper}.
Most existing soliton generation techniques such as power kicking \cite{Brasch:16},  active-capture \cite{yi_active_2016,fpga}, pulse pumping \cite{pulse_herr}, using an auxiliary laser \cite{aux_laser,aux2}, pump laser modulation \cite{Wildi:19}, and fast laser sweeping with a single-sideband modulator (SSB) \cite{ssb} currently involve bulky setups with many off-chip components such as acoustic optic modulators (AOM), circulators, a pulse pump source, or multiple lasers.
Alternative approaches that avoid additional components, including dual-mode pumping \cite{Li:17,weng_dual-mode_2022} and pump laser self-injection locking \cite{Shen2020}, have also been demonstrated, but introduce different practical constraints on variables such as the pump frequency, the accessible soliton detuning range, and the specific resonator designs.
The heater-kick technique \cite{joshi_thermally_2016}, which utilizes simple thermal tuning through metal microheaters deposited above the microresonator, is a well-established and fully integratable technique for soliton generation.
However, the standard kick technique accounts only for fast thermal transients occurring on the scale of tens of microseconds, and the generated soliton comb remains susceptible to instability on longer timescales, since slow thermal instabilities, originating from absorption and self-heating (see Fig. \ref{fig:Intro}a), are enhanced in low-$f_{\text{rep}}$ devices.
Defects or impurities that overlap with the optical mode and, even at small concentrations, can lead to significant absorption \cite{mass_manu,copper}.

In this paper, we present an approach to achieving stable soliton generation through active thermal control of the photonic chip. 
We model and experimentally demonstrate the impact of absorption and the resulting thermal effects on comb dynamics.
We identify several TO instabilities including fast and slow transient instabilities, a steady-state instability causing particular soliton states to become unstable, and a transition-based instability preventing stable soliton switching.
We tailor and implement a control mechanism that overcomes these instabilities despite significantly higher thermal nonlinearities up to 72$\times$ the Kerr nonlinearities and show deterministic soliton generation and strong control over the system.
We accomplish this with fully integratable techniques on low-$f_{\text{rep}}$ devices (16 and 20 GHz) in a compact form factor ($< 1\text{ mm}^2$) and reduce the input optical power requirement for soliton generation by 2.7$\times$ compared to the standard heater-kick technique.
We show that soliton states that survive a few milliseconds with the standard heater-kick technique can be stabilized for days.
Furthermore, we can access soliton states that are practically inaccessible with the heater-kick technique including breather soliton states and high-number soliton states.
Our approach is compatible with automated soliton generation, or turn-key operation, and long-term stability.
We believe this work addresses longstanding challenges of identifying and overcoming the thermal instabilities that impact soliton generation.
 Our results are relevant to all photonic platforms and can be readily implemented for applications such as optical communications, precision metrology, ultralow-noise microwave generation, and LIDAR.

\section{Heat Dynamics and Characterization} \label{sec:heat}
\begin{figure*}[htbp]
	\centering\includegraphics[width=\textwidth]{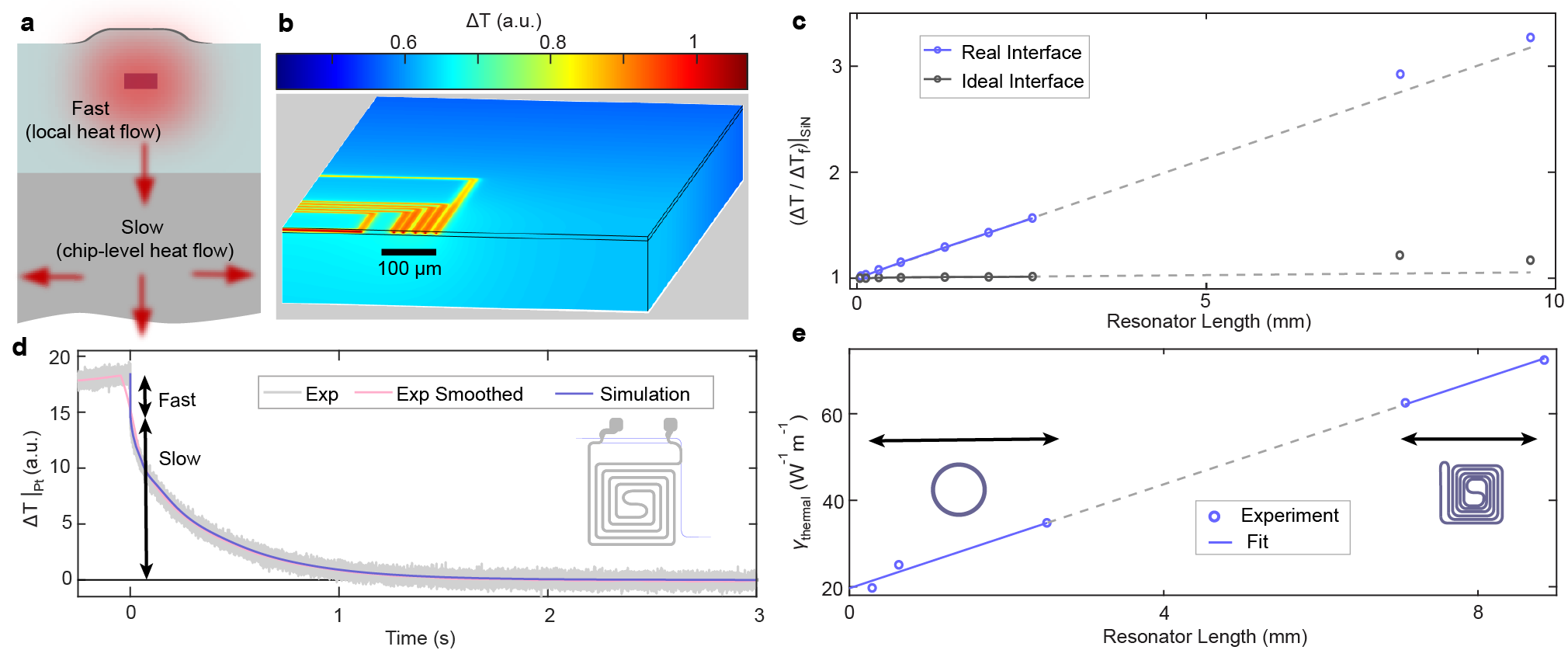}
	\caption{
		\textbf{Thermal characterization of Euler-spiral microresonators. }
		(a) Schematic illustrating the cross-sectional view of a waveguide and self-heating. 
		(b) Steady-state temperature distribution when the spiral microresonator core is the heat source.  
		(c) The simulated normalized change in steady-state temperature of the SiN resonator as a function of resonator length when injecting heat into the spiral microresonator. 
		Small resonator lengths correspond to simple ring resonators while longer lengths correspond to Euler-spiral microresonators. 
		The true interface boundary conditions are compared to an ideal interface. 
		The dashed line is a fit over the ring resonator results. 
		(e) Thermal nonlinear coefficient of SiN microresonators as a function of the resonator length. 
		(d) Experimental and simulated results of the temperature change in a Pt resistance thermometer when a heat source in the waveguide is turned off near t = 0. 
		The simulation and experimental results only match for one value of the heat transfer coefficient. 
	}
	\label{fig:heat}
\end{figure*}

We develop a comprehensive model of heat dynamics in our SiN photonic devices to elucidate the role of absorption and self-heating in the generation and stability of cavity solitons.
We characterize chip-level heat flow experimentally by sweeping a laser's frequency across a cavity resonance and directly measuring the resulting change in on-chip temperature at some distance from the microresonator.
The on-chip temperature is measured by utilizing the thin-film platinum resistor deposited above the microresonator, which can be used as a microheater or as an integrated resistance thermometer \cite{Dacha2026}. 
By applying the appropriate boundary conditions, which account for the non-ideal interface between the bottom of the chip and the top of the chip mount in the simulation model, we see excellent agreement between simulation and experiment, as shown in Fig. \ref{fig:heat}d.
We use our simulation to get the 3D temperature distribution in Fig. \ref{fig:heat}b and the length-normalized steady-state temperature change for several resonator geometries \ref{fig:heat}c.
These results indicate the existence of the following two thermal diffusion processes (see Fig. \ref{fig:heat}a): (1) fast diffusion of heat between the microresonator waveguide and the substrate, and (2) slow diffusion of heat through and out of the chip via the bottom contact with the chip mount.
The steady-state temperature change is given by $\Delta T = \Delta T_f+\Delta T_s$ where $f$ and $s$ denote the fast and slow components, respectively.
The former is a localized phenomenon and only depends on the resonator cross section, and thus is independent of resonator length.
The latter occurs globally across the entire device with a slow spatial decay (see Fig. \ref{fig:heat}b) and carries a strong resonator length dependence (Fig. \ref{fig:heat}c), which poses a hurdle for low-$f_{\text{rep}}$ soliton generation.
We verify these results by measuring the thermal nonlinear coefficient of several microcavities, as shown in Fig. \ref{fig:heat}e, which also shows the linear dependence on resonator length and a high thermal nonlinear coefficient of $\sim$72 W$^{-1}$m$^{-1}$ for the 16-GHz device.
We also investigated thermal paste as a means to improve the chip-to-chip-mount heat transfer, which improved the heat transfer coefficient, but the slow heat diffusion remained non-negligible (see Supplementary Information for details).

\section{Soliton Generation and Thermal Transients}
\label{sec:gen}
Thermal effects complicate soliton generation due to strong thermal nonlinearities which act to push the soliton out of the soliton existence range (SER).
Despite the development of several soliton-generation techniques to overcome such transient thermal effects \cite{Brasch:16,yi_active_2016,pulse_herr,aux_laser,transitions,Li:17,weng_dual-mode_2022,Shen2020,Wildi:19}, few are compatible with full integratability and scalability as off-chip components are required.
The heater-kick technique \cite{joshi_thermally_2016}, which utilizes thermal tuning through metal microheaters deposited above the microresonator, is a well-established and fully integratable technique used for soliton generation.
However, the standard kick technique accounts only for fast thermal transients occurring on the scale of tens of microseconds and does not address instabilities that occur on longer timescales that necessitate a significantly stronger pump for stable operation in the soliton regime.
A technique, known as active capture and stabilization \cite{yi_active_2016,fpga}, that accounts for both timescales has been reported previously, in which a PID is turned on after the generation of the desired soliton state, but it requires the use of external modulators.

\begin{figure*}[htbp]
	\centering\includegraphics[width=\textwidth]{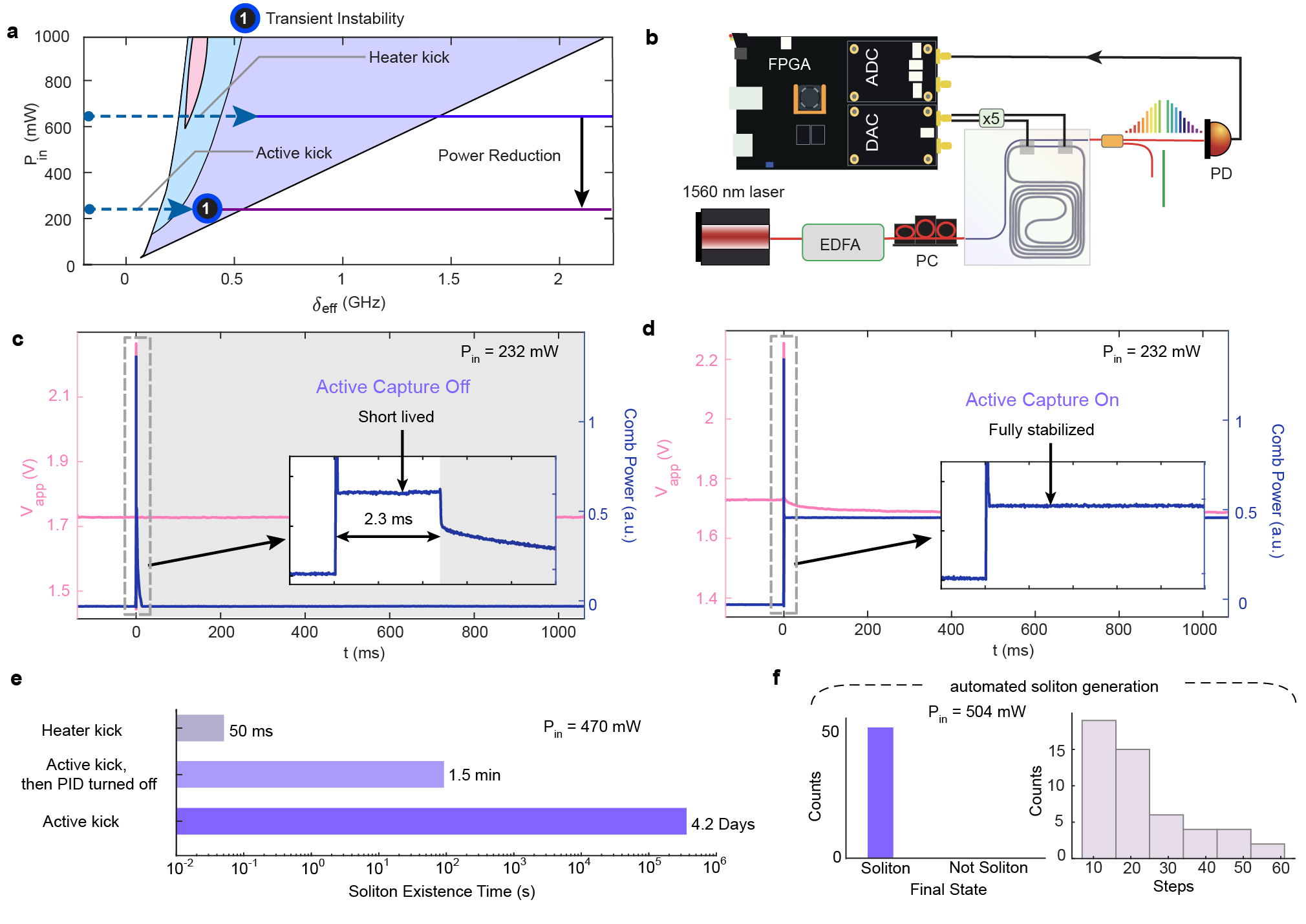}
	\caption{
		\textbf{Overcoming transient temperature effects to get into the soliton regime. }
		(a) Soliton existence map with the minimum powers for both the heater-kick and active-kick techniques.
		The minimum power for the heater-kick technique to reach the soliton regime is 627 mW.
		The minimum power for the active-kick technique to reach the soliton regime is 232 mW.
		(b) Experimental setup for soliton generation. 
		An amplified fixed-frequency laser is sent into the microresonator. 
		The FPGA is used to control the cavity resonance by applying a voltage to the Pt microheaters deposited directly above the resonator. 
		The FPGA uses the detected comb power to do feedback and stabilize the comb. 
		(c, d) Soliton generation with the conventional heater-kick technique (c) and active-kick technique (d) at a pump power of 232 mW.
		For the heater-kick technique, the soliton lasts around 2 ms after the initial kick.
		For the active-kick technique, after a small duration ($\sim100 \:  \mu$s), a PID is turned on with a setpoint based on the current comb power and stabilizes the comb indefinitely.
		(e) Maximum soliton existence time for the heater-kick technique, the active-kick technique, and the active-kick technique where the PID correction is turned off with bias voltage after initial generation.
		With full stabilization, a soliton state can go from a soliton existence time of 50 ms to up to 4.2 days.
		(f) Automated soliton generation results.
		The FPGA runs an algorithm attempting to get into the soliton regime and we check how many times it correctly gets into the soliton regime (left) and the number of steps the FPGA took (right).
	}
	\label{fig:FPGA}
\end{figure*}

Here, we combine the heater-kick \cite{joshi_thermally_2016} and the active-capture techniques \cite{yi_active_2016}; which we term the ``active-kick'' technique.
Our technique relies on an electronic control system, which is implemented using a field-programmable gate array (FPGA) to control the temperature of the microresonator via a voltage applied to the microheater.
The feedback loop stabilizes the comb power and is implemented by filtering out the pump and measuring the comb power using a slow detector; both elements are compatible with full integration.
After a delay of $\sim$50--100 $\mu$s, the FPGA initiates a PID stabilization loop once a soliton state is generated and uses the current comb power as the setpoint.
As shown in Figs. \ref{fig:FPGA}c, d, short-lived soliton states at optical powers of 232 mW, which last only 2.3 ms with the heater-kick technique, can survive quasi-indefinitely with the active-kick technique.

The long-term stability of our scheme implemented on an unpackaged and unshielded device is shown in Fig. \ref{fig:heat}e.
The soliton state, which typically lasts only 50 ms with the heater-kick technique, was stabilized using our technique for $>$ 4 days.
We performed additional testing to assess whether the feedback was only needed for the initial thermal transients or if it was also important to stabilize against ambient temperature fluctuations.
By programming the PID correction to turn off with bias voltage, we observed that the maximum soliton existence time was only 1.5 minutes with many iterations being significantly shorter (see Methods for more details).
Finally, we show that our technique is compatible with automation, such as previous demonstrations \cite{fpga}, via a simple algorithm described in detail in the Methods with the key results shown in Fig. \ref{fig:heat}f.
We ran 50 iterations where in each iteration the FPGA attempts to access a soliton state and stops once a soliton state is generated.
Once the FPGA stops, we verified whether it is indeed a soliton state and how many steps the FPGA took before it stopped (Fig. \ref{fig:heat}f).
Our active-kick technique is able to correctly discriminate whether the cavity output is in the soliton regime or not and access the soliton regime with a 100\% success rate.

\section{Steady-State Instability}
\label{sec:ss}
\begin{figure*}[htbp]
	\centering\includegraphics[width=\textwidth]{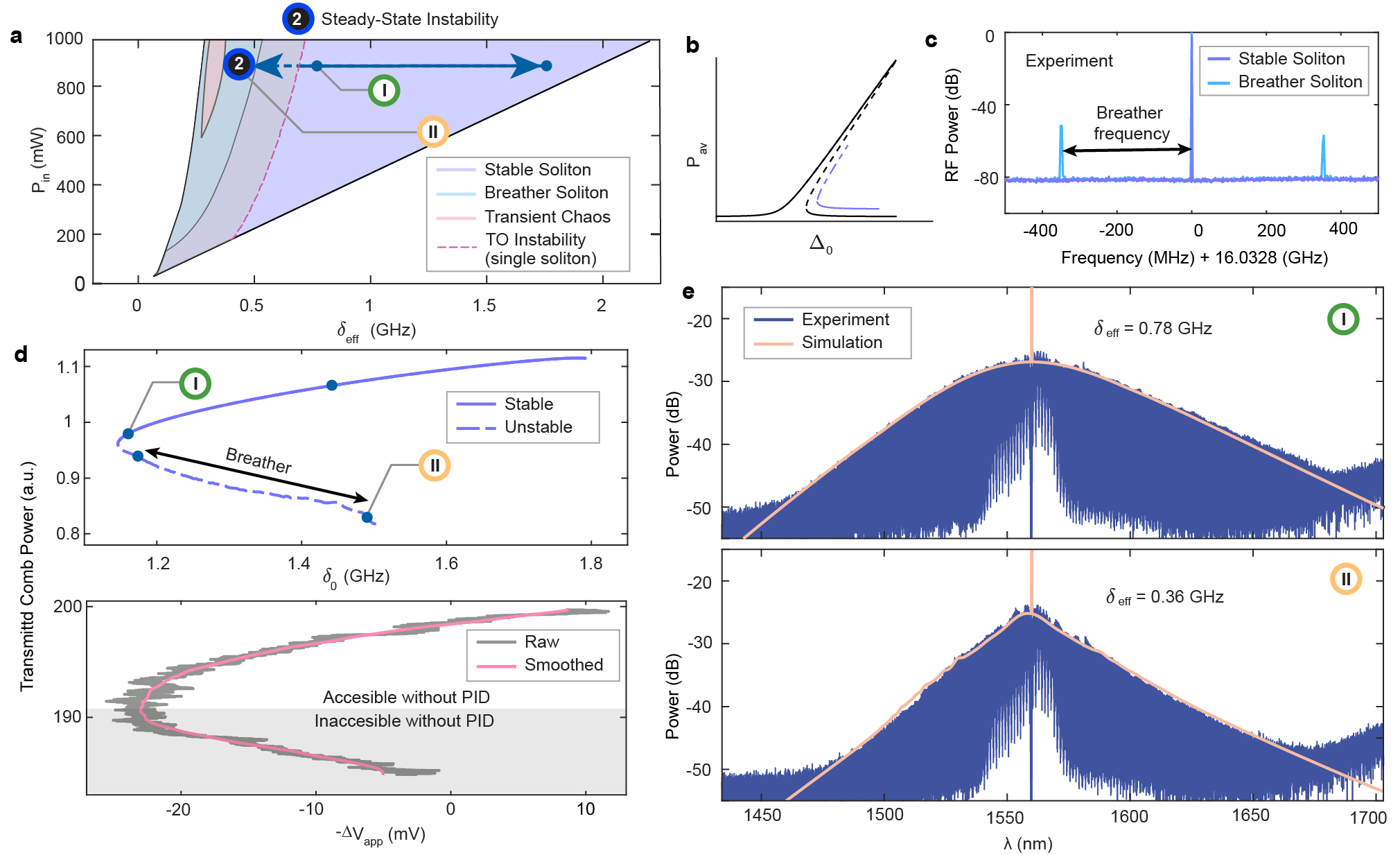}
	\caption{
		\textbf{Steady-state thermo-optic-induced instabilities in soliton states. }
		(a) Soliton existence map with the thermo-optic instability for the single-soliton state.
		The gray region represents soliton states that are inaccessible without comb feedback.
		(b) Illustration of the average power versus cold-cavity detuning for the CW solution (black) and the single-soliton solution (blue) when adding thermo-optic effects.
		Dashed lines indicate unstable solutions.
		(c) Experimental RF spectrum for a stable soliton and breather soliton.
		The breathing frequency in this case is 350 MHz.
		(d) TO instability in the single-soliton state for simulation (top) and experiment (bottom).
		The simulation shows soliton solutions which become unstable when the thermo-optic term is added.
		The breather soliton lies along the unstable solutions.
		(e) Experimental and simulated single-soliton spectrum for a stable soliton (top) and breather soliton (bottom).
	}
	\label{fig:SS}
\end{figure*}

We now examine instabilities in the soliton regime in the presence of the thermo-optic term in the cavity dynamics model.
While the steady-state solutions of the Kerr dynamics are functions of the effective detuning $\delta_{\text{eff}}$, defined without considering thermo-optic effects, the steady-state solutions of the combined system are functions of the cold-cavity detuning $\delta_0$.
These two quantities differ by the non-zero thermal nonlinearity and are related by the average intracavity power (see Methods eq. 2 for the relation). 
For a positive thermo-optic coefficient, if the average power decreases versus the effective detuning, it is possible for the steady-state solutions to fold such that there are multiple solutions versus cold cavity detuning with the upper branch being unstable and the lower branch being stable (illustrated in Fig. \ref{fig:SS}b).
The presence of the thermo-optic term results in the single-soliton state becoming unstable at small detunings due to the rapid decrease in intracavity power versus effectice detuning (as shown in Fig. \ref{fig:SS}).
We observe this instability experimentally, as shown in the lower panel of Fig. \ref{fig:SS}d, where for a range of heater voltages, or cold cavity detunings, there are two soliton solutions.
The results agree with the simulation shown in the upper panel of Fig. \ref{fig:SS}d.
If the cold cavity detuning is decreased to the edge of the stable branch in the absence of feedback, an effective detuning corresponding to an unstable soliton state is eventually reached.
At this point, the effective detuning rapidly decreases on a timescale corresponding to the instability and, depending on the power and the initial state, the intracavity field can settle either into a chaotic state or pass through a soliton transition and land in a new stable soliton state.

Despite this instability, it is possible to achieve small effective detunings, including breather soliton states \cite{breath1,breath2,breath3}, which lie in the unstable region (Fig. \ref{fig:SS}a).
We show the optical spectra of a stable single soliton in the top panel of Fig. \ref{fig:SS}e, which features a characteristic sech$^2$ envelope along with the RF spectrum in Fig. \ref{fig:SS}c and has a low-noise RF tone in Fig. \ref{fig:SS}c.
We also show results for a single breather soliton which are characterized by a breathing frequency, as seen by the new RF tone in the RF spectrum in Fig. \ref{fig:SS}c and have a triangular shaped optical spectral envelope \cite{breath1,breath3}, as shown in the lower panel of Fig. \ref{fig:SS}e.
The experimental optical spectra match the simulated spectra, and the breathing frequency, which in this case is approximately 350 MHz,  corresponds well with other breather soliton demonstrations \cite{breath1,breath2,breath3}.
In Fig. \ref{fig:Transitions}b, we show the simulated and experimentally measured soliton existence ranges for several multi-soliton states.
Similar to the single-soliton state, there are stable and unstable branches.
Compared to the single-soliton state, multi-soliton states are more thermally stable due to a larger change in average power versus effective detuning.
This results in an increased amount of stable soliton states and the same effective detuning range being supported by a larger range of cold cavity detunings often referred to as ``self-locking.''
We note that our technique allows us to access high soliton number states that are practically inaccessible with the heater-kick technique since they are short-lived without immediate feedback.
As shown in Fig. \ref{fig:Transitions}, soliton numbers as high as $\sim$50 can be accessed, whereas we are limited to $\sim$9 with the standard heater-kick method.

\section{Controlled Soliton Transitions}\label{sec:transitions}
\begin{figure*}[htbp]
	\centering\includegraphics[width=\textwidth]{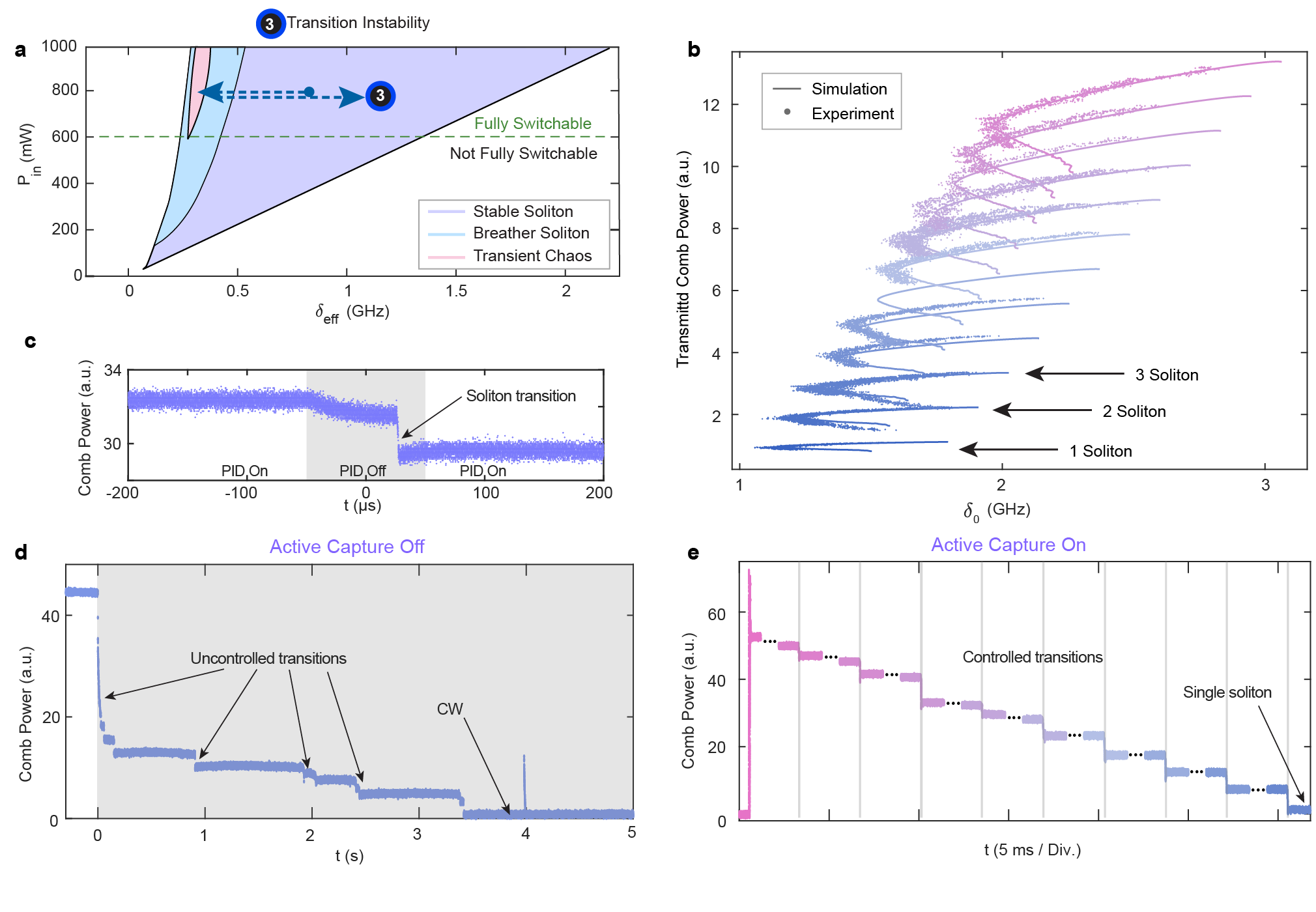}
	\caption{
		\textbf{Controlled soliton transitions and soliton existence map. }
		(a) Soliton existence map showcasing soliton transitions.
		(b) Experimental and simulated soliton existence for a power of 900 mW for several soliton numbers.
		The experimental data is scaled vertically to match simulation, and each soliton curve is manually moved horizontally to account for drift.
		(c)
		Example soliton transition. 
		The PID correction is turned off with bias voltage during the time period indicated by gray and the voltage is stepped to induce a change in detuning which then causes a soliton transition. 
		When the PID is turned back on, it uses the new comb power as the new set point for the PID and stabilizes the state indefinitely. 
		(d) Uncontrolled soliton transitions with a simple detuning ramp.
		An active-kick is initiated, and the PID correction is disabled without bias voltage, while starting a ramp of the applied voltage to switch the soliton number.
		(e) Repeated controlled soliton transitions. 
		An active-kick is initiated and a controlled soliton transition is performed to lower the soliton number down to the single-soliton state.
		A delay time exists between each color trace (not shown). 
	}
	\label{fig:Transitions}
\end{figure*}

Although high-number soliton states tend to have increased TO stability and self-locking, higher conversion efficiencies, and higher RF powers, there is still interest in single-soliton states due to their predictable optical spectra.
For cavity dynamics governed by the Lugiato-Lefever equation \cite{LLE} with no higher-order dispersion terms and only Kerr nonlinearity, the number of pulses and the positioning of the pulses are random, which leads to a degree of randomness in the optical spectra.
In our system, the initially generated soliton states are also high-number soliton states, which means getting to the single-soliton requires several multi-soliton transitions.
The standard way to do these transitions is to decrease the cold cavity detuning until the far-left boundary is reached \cite{transitions} as shown in Fig. \ref{fig:Transitions}a.
We first try this technique in Fig. \ref{fig:Transitions}e where the PID correction is disabled without bias voltage after an active-kick and  then we slowly ramp the applied heater voltage.
We observe that there is an immediate cascading of soliton states which is hard to control, followed by longer and random delays between subsequent transitions.
The initial cascade arises from the sudden jump in applied voltage, analogous to an uncompensated heater kick.
The later irregular transitions are attributed to the enhanced TO effect in this device and to ambient fluctuations.

By applying the same central idea from the active-kick technique of providing feedback faster than the timescale of the instability, we show significantly more controlled soliton transitions in Fig. \ref{fig:Transitions}c-d.
Our technique consists of recapturing a soliton state where we turn the PID correction off with bias voltage, programmatically change the applied heater voltage, then quickly turn the PID corrections back on after a short delay of $\sim$100 $\mu$s with the current comb power as the set point (Fig. \ref{fig:Transitions}c).
This naturally accommodates the change in soliton number, which features a sudden and unpredictable jump in comb power while minimizing the downtime where the soliton state is unstabilized and the resulting ambient fluctuations.
Furthermore, we can stabilize the new state provided it lasts longer than 100 $\mu$s, as illustrated in Fig. \ref{fig:Transitions}d where each color represents a different controlled soliton transition.
The experiment consists of applying the technique described above to change the detuning and recapture the new soliton state repeatedly.
We monitor the transitions by recording the comb power on a real-time oscilloscope (Fig. \ref{fig:Transitions}d), and we verify the soliton state by simultaneously measuring the optical and RF spectrum.
We observe controlled transitions from where the active-kick initially accesses the soliton regime, down to the single-soliton state.

\section{Conclusion and outlook} \label{sec:conc}
We demonstrate an approach to deterministic Kerr soliton generation even in devices with high thermal nonlinearities and in highly compact resonator geometries.
Through comprehensive modeling of the heat dynamics of the photonic chip, we determine how absorption and self-heating play a role in the instabilities that prevent soliton generation, especially in low-FSR resonators and demonstrate a technique to overcome them. 
We identify multiple instability mechanisms including a transient instability while accessing the soliton regime, steady-state instability within a soliton state, and a transition-based instability to access other soliton states.
With active thermal control, we show it is possible to overcome all of these instabilities and achieve strong control over the comb and extend short-lived soliton states by several orders of magnitude, even for low-FSR, (16 GHz) spiral devices with compact 1 mm$^2$ footprints.
Our technique can be applied as a general tool for deterministic soliton generation in arbitrary resonator designs and across all photonic platforms. 
We believe this work addresses a longstanding challenge of arbitrary deterministic soliton generation which can be immediately implemented for several applications such as optical communications, precision metrology, ultralow-noise microwave generation, and LIDAR.

\section{Acknowledgements}
This work was performed in part at the Cornell Nano-Scale Facility, which is a member of the National Nanotechnology Infrastructure Network, supported by the NSF.
This work was supported by Defense Advanced Research Projects Agency of the US Department of Defense (Grant No. HR0011-24-2-0371), Army Research Office (ARO) (Grant No. W911NF-24-2-0204), and National Science Foundation (NSF) (PHY-2513522).

\section{Author Contributions}
G.J.B., Y.Z., and A.L.G. conceived the idea for the project.
G.J.B. performed the experiments with assistance from S.K.D, Y.Z., and J.E.
G.J.B. performed the data analysis with input from all authors.
G.J.B. and Y.Z. designed the devices.
K.J.M. fabricated the devices under the supervision of M.L.
G.J.B., S.K.D. and A.L.G. wrote the manuscript with feedback from all
authors. 
M.L. and A.L.G. supervised the project.

\section{Competing Interests}
The authors declare no conflicts of interest at this time.

\section{Methods} \label{sec:methods}
\setcounter{figure}{0}
\renewcommand{\figurename}{Extended Fig.}

\subsection{Design and characterization of Euler-spiral microresonators}
\begin{figure*}[htbp]
	\centering\includegraphics[width=\textwidth]{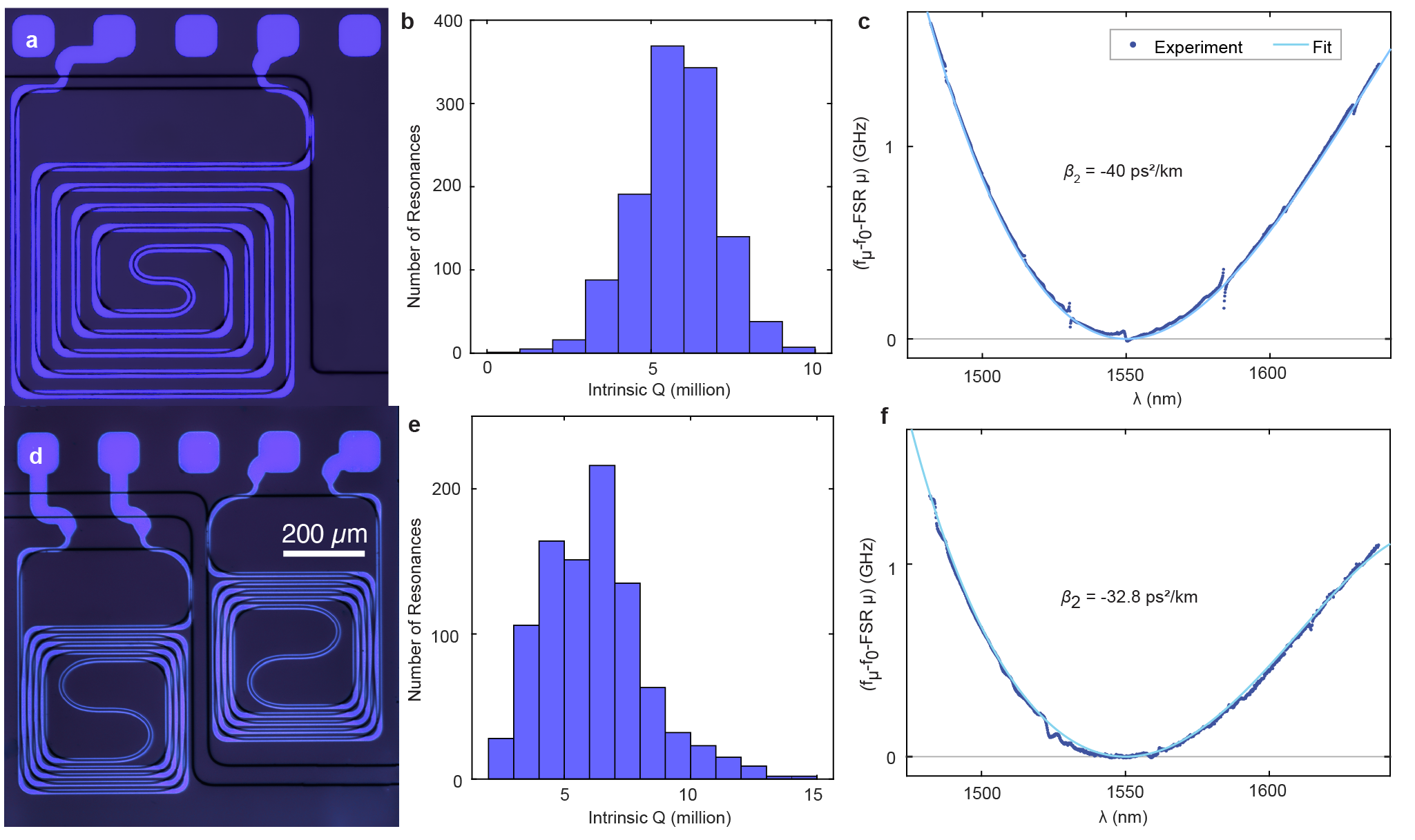}
	\caption{
		\textbf{Optical characterization of Euler-spiral microresonators. }
		(a), (d) Microscope pictures of Euler-spiral microresonators. 
		(a) 16 GHz Euler-spiral with 50 $\mu$m waveguide spacing. 
		(d) 20 GHz Euler-spiral with 20 $\mu$m waveguide spacing. 
		The other plots are shown for the left device. 
		(b), (e) Histogram of measured intrinsic quality factors of the devices. 
		(c), (f) Dispersion operator of the devices.
	}
	\label{fig:Q}
\end{figure*}

We designed and fabricated several Euler-spiral microresonators composed exclusively of Euler-style `r' and `u' bends, connected with straight waveguide sections.
Example microscope pictures of fabricated devices are shown in Extended Fig. \ref{fig:Q}.
At the spiral center, two `u' bends connect the clockwise and counter-clockwise waveguide paths. 
The main body of the spiral is comprised of alternating `r' bends and straight sections, and it is closed with two adjacent `r' bends at the outer edge.
Similar to an Archimedes spiral, we design the spacing between adjacent arms of the spiral to be fixed with a minimum waveguide-to-waveguide spacing of 50 $\mu$m for Extended Fig. \ref{fig:Q}a and 20 $\mu$m for Extended Fig. \ref{fig:Q}b.
The FSR of the microresonators are around 16 and 20 GHz for Extended Fig. \ref{fig:Q}a and b accordingly. 
The top device in Extended Fig. \ref{fig:Q} fits within a 1 x 1 mm$^2$ field and the bottom devices fit within a 0.6 x 0.6 mm$^2$ field.
The waveguide heights are $\sim$690 nm, and the widths are $\sim$1500 nm.
This width is relatively narrow which sacrifices the maximally achievable Q, which tends to increase with width \cite{Ji:17}, in order to minimize coupling to higher-order modes (HOM) \cite{chen_general_2012}.

The devices are characterized via a standard dispersion measurement \cite{disp1,Yi:15} where a tunable laser is swept quickly across a wavelength range extending beyond the C-band.
The time axis is calibrated using a fiber-based Mach-Zehnder interferometer (MZI) as a reference where the fiber MZI was calibrated by measuring the dispersion with a fully stabilized optical frequency comb \cite{Yi:15}.
The results of the dispersion measurement are shown in Fig. \ref{fig:Q} for the TE\textsubscript{00} mode. 
As seen in Ext. Fig. \ref{fig:Q}b and Ext. Fig. \ref{fig:Q}e, the intrinsic $Q$ of the devices across 1480 to 1640 nm is 5.8 $\pm$ 1.3  and 6.2 $\pm$ 2  million accordingly.
Absolute input optical power requirements can be lowered by further improving the quality factor of the resonators through advanced fabrication techniques such as chemical mechanical polishing (CMP) and multi-pass deposition \cite{Ji:17}.
The dispersion operator is shown in Fig. \ref{fig:Q}c and Fig. \ref{fig:Q}f.
The devices have anomalous dispersion with minimal mode-splittings, and any observed splittings have low coupling strength.
The negligible splitting strength is also illustrated in the single-soliton optical spectrum for the 16 GHz device in Fig. \ref{fig:Intro}d.

We fabricate the devices on 100 mm silicon wafers with a 4 $\mu$m wet thermal oxide buffer. 
A 690 nm thick SiN layer is deposited via low-pressure chemical vapor deposition (LPCVD) in two steps, with high-temperature annealing between depositions to prevent cracking. 
An oxide hardmask is deposited atop the SiN, and the waveguides are patterned using electron-beam lithography and etched with a plasma dry etch. 
After etching, the waveguides undergo a 3-hour anneal in an argon environment and are clad with high-temperature silicon dioxide. 
Platinum microheaters are defined on the cladding oxide through a lift off process, and we perform a final deep silicon etch prior to dicing to enable low-loss edge couplers.

\subsection{Resistance thermometer measurement}
For the resistance thermometer measurements shown in Fig. \ref{fig:heat}d,  the temporal response is characterized experimentally by scanning an amplified laser across a resonance of the microresonator and monitoring the temperature of a platinum integrated resistance thermometer which is situated directly above the resonator.
Note that the temperature sensor is spatially separated from the heat source, so the resulting impulse response is a result of both the cavity temperature evolution as well as the heat transfer from the resonator to the platinum thermometer.
The platinum microheater, which is the same structure typically used to control the resonance frequency of the device, is used here to sense the temperature by utilizing its temperature-dependent resistance.
Recently, we applied this technique to stabilize the absolute frequency of a microresonator \cite{Dacha2025}. 
The laser is scanned slowly and with sufficient power to observe a triangular resonance shape where the sharp edge is utilized as a step-off response for the thermal characterization.
We also perform COMSOL Multiphysics simulations to find the temperature evolution of the platinum after injecting heat into the SiN microresonator and find agreement with the experimental data for a heat transfer rate of $1.9 \times 10^{3} \text{ W}\text{m}^{-1}\text{K}^{-1}$.
More details on the simulation can be found in the Supplementary Information.

\subsection{Sum measurement}
To measure the thermal nonlinear coefficients in Fig. \ref{fig:heat}e, we employ the sum measurement technique described in ref. \cite{rokhsari_loss_2004}.
First, we perform linear characterization of the device by carrying out a $Q$ measurement at low optical input power.
Then, we increase the input power until we observe a strongly tilted resonance and scan a laser slowly ($\sim$10 mHz) via piezo tuning across a resonance of the microresonator. 
We ensure that the input power remains below the OPO (optical parametric oscillator) threshold.
We convert the applied voltage to frequency via a wavemeter calibration.
Using the linear parameters of the resonance from the $Q$ measurement coupled with the numerical model described above, we extract the total nonlinear coefficient of the device.
We determine the optical nonlinear coefficient by simulating the effective area of the optical mode and by assuming $n_2 = 2.5 \times 10^{-19} \text{ m}^{2} \text{ W}^{-1}$.
The optical nonlinear coefficient can then be calculated via $\gamma_{\text{opt}} = \frac{2\pi n_2}{\lambda_0 A_{\text{eff}}}$ and is approximately 1.05 W$^{-1}$ m$^{-1}$ for our devices. 
We assume there are no other significant nonlinear terms, and the thermal nonlinear coefficient is extracted as $\gamma_{\text{thermal}} = \gamma_{\text{total}} - \gamma_{\text{optical}}$.

\subsection{Automated Soliton Generation}
\begin{figure*}[htbp]
	\centering\includegraphics[width=\textwidth]{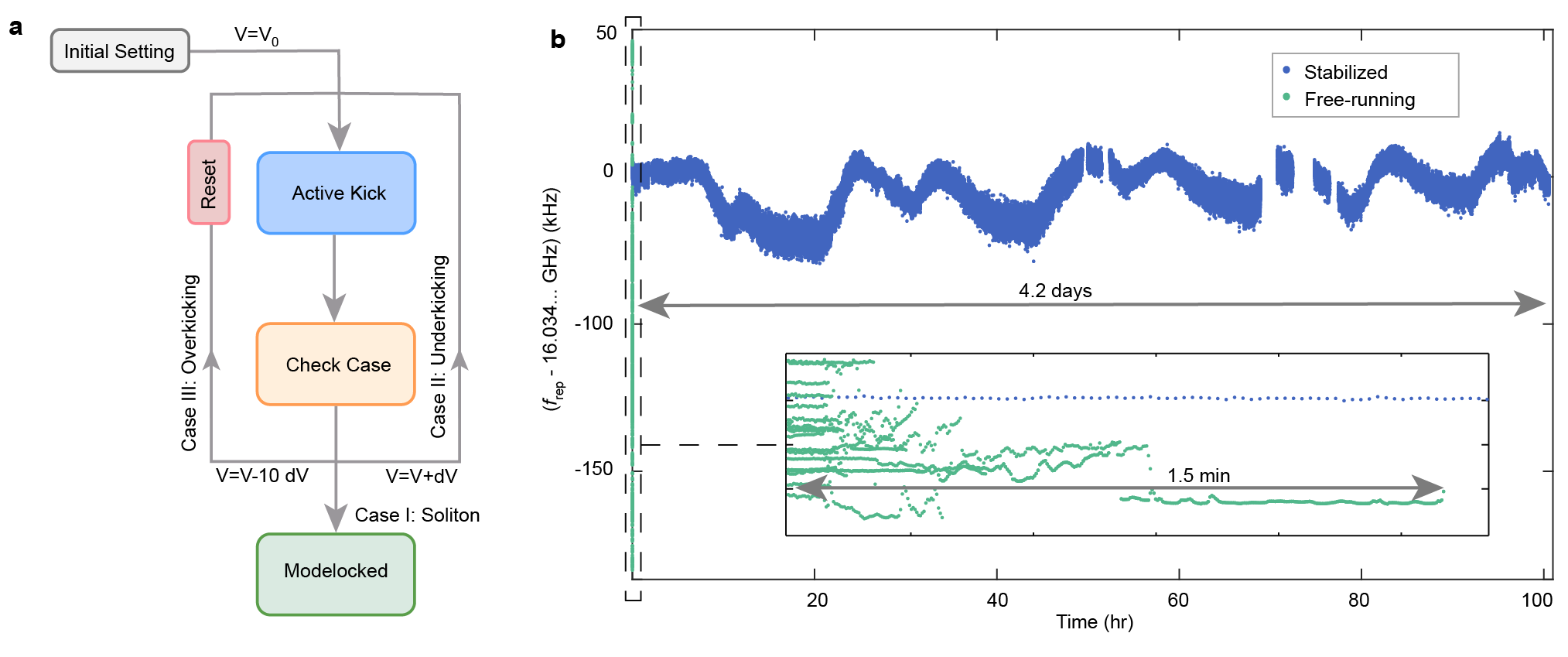}
	\caption{
		\textbf{Automated Kerr comb generation and long-term stability.}
		(a) Illustration of automated soliton generation algorithm. The loop is repeated until a soliton state is generated.
		(b) Long term stability of the generated soliton microwave frequency. 
		White regions  indicate times where data was not taken.
		We show the free-running case where the soliton is generated with the active-kick technique and then we turn the PID correction off with bias voltage. The maximum soliton existence time is 1.5 minutes and often it goes out of the soliton regime earlier. 
		We also show the stabilized soliton state which has a soliton existence time of 4.2 days. 
	}
	\label{fig:autolong}
\end{figure*}

We show that it is possible to automate the soliton generation process for our 16 GHz Euler-spiral microresonator via microheater control with a straightforward algorithm as illustrated in Ext. Fig. \ref{fig:autolong}a.
An FPGA is used for the control electronics and is interfaced with a laptop running the automation algorithm.
We use a 5$\times$ voltage amplifier after the FPGA DAC and before the microheaters, and the detected comb power from a slow photodetector is fed back to the FPGA ADC.
We start with an initial setting such that the laser is off-resonance and the applied voltage is low (red-detuned) and the loop begins.
In each iteration, the loops start with an active-kick where the FPGA first implements the standard heater kick (see the Supplementary Information for an example heater kick).
If the comb power after some delay time ($\sim$50 $\mu$s) falls within a certain range, the PID will turn on with the current comb power as the set point.
Next, the algorithm checks to determine whether a soliton is generated and what case the system falls into.
We do this by looking at the comb power, the change in comb power, and the RMS fluctuation of the comb power.
If a soliton is not generated (case I), then the applied voltage is either too low (case II: underkicking) or too high (case III: overkicking).
In the case of underkicking, the sweep goes all the way through resonance past the soliton state and to an off-resonance CW state.
The voltage can be consecutively increased until the kick is aligned with the soliton step and a modelocked soliton state can be achieved.
However, even small misalignments of the kick to the soliton step---for example through ambient temperature fluctuations, drift, etc.---can prohibit stable soliton generation.
If the voltage becomes too high, which we call overkicking, the kick can come before the soliton step and keep the system in resonance usually in a chaotic state.
In this case, a reset is needed where the voltage is changed significantly to make sure the laser is tuned fully off resonance before the next kick.
Since it is easier to approach from the underkicking side, we weight the change in voltage for the case of overkicking to be 10 times higher than underkicking.
The algorithm ends when it believes a soliton has been generated and we verify the existence of a soliton state due to a sharply peaked RF power which only occurs in the soliton regime (note that we do not observe perfect soliton crystal states or PSCs).

\subsection{Numerical simulations}
In order to model the optical nonlinear dynamics of the cavity, we use the Lugiato-Lefever equation \cite{LLE} with higher-order dispersion and a frequency-dependent coupling coefficient:
\begin{align*}
	t_R \frac{\partial E(t,\tau)}{\partial t} &= 
	\Big[-\mathscr{F}^{-1}\{\alpha(f)\} - i\delta_{\text{eff}} + i\gamma L|E|^2 +\\ &iL
	\sum_{k\geq2}\frac{\beta_k}{k} \left(i \frac{\partial}{\partial \tau}\right)^k
	\Big]E(t,\tau) + \mathscr{F}^{-1}\Big\{\sqrt{\theta(f)}\Big\}E_{\text{in}}
\end{align*}
Where $t_R$ is the cavity roundtrip time, $\alpha(f) = [\alpha_i + \theta(f)]/2$ is the total field loss in one round trip, $\gamma$ is the optical nonlinear coefficient of the cavity, $L$ is the cavity length, $\theta$ is the coupling coefficient, $\beta_k$ is the $k$th order dispersion coefficient, and $\delta$ is the pump detuning to its nearest resonance.
We define the detuning of the Kerr dynamics, without considering thermo-optic effects, as the effective detuning or $\delta_{\text{eff}}$, whereas we define the detuning of the combined system as the cold cavity detuning, $\delta_0$.
These two quantities are related via the average intracavity power (see Methods eq. \ref{ssto} for the relation).
For the soliton existence map, we neglect the frequency dependent coupling and higher-order dispersion and we use the following parameters: $\alpha = 0.01521$, $\theta = 0.01856$, $\gamma = 1$ W$^{-1} $ m$^{-1}$, $L = 8.8757$ mm, $\beta_2 = -40$ ps$^2$ km$^{-1}$, and FSR $ = 1/t_R = 16.0367$ GHz.
We use a Newton-Raphson solver for the stable soliton solutions \cite{Qi:19} and a split-step method for the breather soliton and transient chaos regime.
The transient chaos region is defined by the region where seeding a single-soliton solution ends in decay to the CW solution.
For the optical spectra shown in Fig. \ref{fig:SS}, the simulation is averaged over many round trips.
More details on simulating the optical nonlinear dynamics including the parameters used for higher-order dispersion and frequency-dependent coupling can be found in the Supplementary Information.

For the simulation of multi-soliton states shown in Fig. \ref{fig:Transitions}b, we use an approximation which takes advantage of the localized nature of the solitons to quickly solve for an $N$ soliton state.
We solve for 2 different soliton number solutions, then we can approximate the $N$ soliton state by a linear combination of these solutions.
We have checked for particular values that this approximation closely matches simulated numerical solution results.
However, we observe that breather soliton solutions close to the edge of the existence range can have larger deviations which gives rise to the ripples on the edge of the soliton states on the lower branches in Fig. \ref{fig:Transitions}b.

For the simulations of the comb power versus detuning in Fig. \ref{fig:SS} and Fig. \ref{fig:Transitions} as well as the thermo-optic instability in Fig. \ref{fig:SS}a, we also use an additional thermo-optic term to add the heat dynamics.
In this case, we do not have to consider the time dependence of the heat as we are interested in quasi-steady-state solutions, or solutions which do not change with respect to  the fastest TO timescale.
This simplifies the analysis since steady-state solutions with LLE are simply shifted solutions in the combined system.
While stable solitons are true steady-state solutions, breather solitons are not and this approximation must be justified.
However, breather solitons oscillate faster than 10 ns and the fastest TO timescale is on the order of 10 $\mu$s, so the heat dynamics should only see the averaged power of the breather over many round trips.
Therefore, we believe this is a good assumption, and the solutions are simply shifted as:
\begin{align}
	\label{ssto}
	E(\delta_0) &= E(\delta_{\text{eff}}) \\
	\delta_0 &= \delta_{\text{eff}} + \gamma_{\text{thermal}} L P_{\text{av}}
\end{align}
Where $\gamma_{\text{thermal}}$ is the thermal nonlinearity and we use a value of 72.5 W$^{-1}$ m$^{-1}$ for our 16 GHz device.
For a positive TO coefficient, if the average power decreases, then the soliton solutions can become multivalued and the upper branch becomes unstable.
Below a certain input power, like in Fig. \ref{fig:SS}a, it is possible for a soliton state to be completely unstable.
For the soliton existence map in Fig. \ref{fig:SS}a, we take the left most value of the upper branch as the beginning of the TO instability.
We have also verified for certain input powers that we get the same boundary when we simulate dynamically.
We did this with (1) the full combined LLE + TO system where we make the thermo-optic timescale artificially fast and (2) by dynamically evolving the temperature dynamics while assuming the intracavity field is in its LLE quasi–steady state. 
Please see the Supplementary Information for more details on stability analysis.

\subsection{Multi-Soliton Existence}
For the single-soliton existence shown in Fig. \ref{fig:SS}d, we accessed the single-soliton state then varied the set point of the PID.
We adjusted the set point by changing the input signal to the PID which was after a summing amplifier which combined the comb power with the output of an AWG.
For the AWG settings, we used a triangular waveform scanned slowly at $\sim$5 Hz, which is significantly slower than the bandwidth of the feedback.
We recorded the data on a real-time oscilloscope and averaged over many traces.

For the multi-soliton existence range measurements shown in Fig. \ref{fig:Transitions}b,  we accessed different soliton states and recorded the comb power and applied voltage on a National Instruments DAC while adjusting the setpoint to an extremum of the soliton state.
We focused on going past the stability boundary then to the extrema associated with higher comb powers.
We then repeated this for many different soliton numbers.
For the 2 and 3 soliton state, we took repeat data where we went to the other extrema associated with lower comb power.
For the 5-soliton state, we happen to have gone slightly further into the lower comb power branch than the other states.
For the analysis, we converted the applied voltage to an electrical power via an I-V relation and that power to the frequency shift via a previous heater calibration (see Supplementary Information for details).
Since this measurement was performed over a long period of time, drift is significant and will alter the applied voltage and shift the solitons artificially.
Therefore, for each soliton state, we shifted the total frequency shift of the experimental data to align the value for the smallest soliton detuning to the simulated results.
Finally, we scale the entire experimental data, including all of the soliton states, by a y-scale stretching factor in the comb power to better match simulation.

\section{References}
\bibliographystyle{apsrev4-1}
\bibliography{bib.bib}

@article{Ji:17,
author = {Xingchen Ji and Felippe A. S. Barbosa and Samantha P. Roberts and Avik Dutt and Jaime Cardenas and Yoshitomo Okawachi and Alex Bryant and Alexander L. Gaeta and Michal Lipson},
journal = {Optica},
number = {6},
pages = {619--624},
publisher = {Optica Publishing Group},
title = {Ultra-low-loss on-chip resonators with sub-milliwatt parametric oscillation threshold},
volume = {4},
month = {Jun},
year = {2017},
url = {https://opg.optica.org/optica/abstract.cfm?URI=optica-4-6-619},
doi = {10.1364/OPTICA.4.000619},
}

@article{mass_manu,
author = {Xinru Ji and Rui Ning Wang and Yang Liu and Johann Riemensberger and Zheru Qiu and Tobias J. Kippenberg},
journal = {Optica},
number = {10},
pages = {1397--1407},
publisher = {Optica Publishing Group},
title = {Efficient mass manufacturing of high-density, ultra-low-loss Si3N4 photonic integrated circuits},
volume = {11},
month = {Oct},
year = {2024},
url = {https://opg.optica.org/optica/abstract.cfm?URI=optica-11-10-1397},
doi = {10.1364/OPTICA.529673},
}

@Article{copper,
author={Ji, Xinru
and Li, Xurong
and Qiu, Zheru
and Wang, Rui Ning
and Divall, Marta
and Gelash, Andrey
and Lihachev, Grigory
and Kippenberg, Tobias J.},
title={Deterministic soliton microcombs in Cu-free photonic integrated circuits},
journal={Nature},
year={2025},
month={Oct},
day={01},
volume={646},
number={8086},
pages={843-849},
issn={1476-4687},
doi={10.1038/s41586-025-09598-4},
url={https://doi.org/10.1038/s41586-025-09598-4}
}

@Article{disp1,
author={Del'Haye, P.
and Arcizet, O.
and Gorodetsky, M. L.
and Holzwarth, R.
and Kippenberg, T. J.},
title={Frequency comb assisted diode laser spectroscopy for measurement of microcavity dispersion},
journal={Nature Photonics},
year={2009},
month={Sep},
day={01},
volume={3},
number={9},
pages={529-533},
issn={1749-4893},
doi={10.1038/nphoton.2009.138},
url={https://doi.org/10.1038/nphoton.2009.138}
}

@Article{Dacha2025,
	author={Dacha, Sai Kanth
	and Zhao, Yun
	and McNulty, Karl J.
	and Bhatt, Gaurang R.
	and Lipson, Michal
	and Gaeta, Alexander L.},
	title={Frequency-stable nanophotonic microcavities via integrated thermometry},
	journal={Nature Photonics},
	year={2025},
	month={Nov},
	day={03},
	issn={1749-4893},
	doi={10.1038/s41566-025-01789-9},
	url={https://doi.org/10.1038/s41566-025-01789-9}
}

@Article{Dacha2026,
	author={Dacha, Sai Kanth
	and Zhao, Yun
	and McNulty, Karl J.
	and Bhatt, Gaurang R.
	and Lipson, Michal
	and Gaeta, Alexander L.},
	title={Frequency-stable nanophotonic microcavities via integrated thermometry},
	journal={Nature Photonics},
	year={2026},
	month={Jan},
	day={01},
	volume={20},
	number={1},
	pages={71-78},
	issn={1749-4893},
	doi={10.1038/s41566-025-01789-9},
	url={https://doi.org/10.1038/s41566-025-01789-9}
}

@Article{transitions,
author={Guo, H.
and Karpov, M.
and Lucas, E.
and Kordts, A.
and Pfeiffer, M. H. P.
and Brasch, V.
and Lihachev, G.
and Lobanov, V. E.
and Gorodetsky, M. L.
and Kippenberg, T. J.},
title={Universal dynamics and deterministic switching of dissipative Kerr solitons in optical microresonators},
journal={Nature Physics},
year={2017},
month={Jan},
day={01},
volume={13},
number={1},
pages={94-102},
issn={1745-2481},
doi={10.1038/nphys3893},
url={https://doi.org/10.1038/nphys3893}
}

@article{ye_integrated_2022,
	title = {Integrated, {Ultra}-{Compact} {High}-{Q} {Silicon} {Nitride} {Microresonators} for {Low}-{Repetition}-{Rate} {Soliton} {Microcombs}},
	volume = {16},
	issn = {1863-8899},
	url = {https://onlinelibrary.wiley.com/doi/abs/10.1002/lpor.202100147},
	doi = {10.1002/lpor.202100147},
	language = {en},
	number = {3},
	urldate = {2023-05-22},
	journal = {Laser \& Photonics Reviews},
	author = {Ye, Zhichao and Lei, Fuchuan and Twayana, Krishna and Girardi, Marcello and Andrekson, Peter A. and Torres-Company, Victor},
	year = {2022},
	note = {\_eprint: https://onlinelibrary.wiley.com/doi/pdf/10.1002/lpor.202100147},
	pages = {2100147},
}

@article{liu_photonic_2020,
	title = {Photonic microwave generation in the {X}- and {K}-band using integrated soliton microcombs},
	volume = {14},
	copyright = {2020 The Author(s), under exclusive licence to Springer Nature Limited},
	issn = {1749-4893},
	url = {https://www.nature.com/articles/s41566-020-0617-x},
	doi = {10.1038/s41566-020-0617-x},
	language = {en},
	number = {8},
	urldate = {2022-11-09},
	journal = {Nature Photonics},
	author = {Liu, Junqiu and Lucas, Erwan and Raja, Arslan S. and He, Jijun and Riemensberger, Johann and Wang, Rui Ning and Karpov, Maxim and Guo, Hairun and Bouchand, Romain and Kippenberg, Tobias J.},
	month = aug,
	year = {2020},
	note = {Number: 8
Publisher: Nature Publishing Group},
	pages = {486--491},
}

@article{yang_bridging_2018,
	title = {Bridging ultrahigh-{Q} devices and photonic circuits},
	volume = {12},
	copyright = {2018 The Author(s)},
	issn = {1749-4893},
	url = {https://www.nature.com/articles/s41566-018-0132-5},
	doi = {10.1038/s41566-018-0132-5},
	language = {en},
	number = {5},
	urldate = {2022-11-28},
	journal = {Nature Photonics},
	author = {Yang, Ki Youl and Oh, Dong Yoon and Lee, Seung Hoon and Yang, Qi-Fan and Yi, Xu and Shen, Boqiang and Wang, Heming and Vahala, Kerry},
	month = may,
	year = {2018},
	note = {Number: 5
Publisher: Nature Publishing Group},
	pages = {297--302},
}

@article{Liu:18,
author = {Guangyao Liu and Vladimir S. Ilchenko and Tiehui Su and Yi-Chun Ling and Shaoqi Feng and Kuanping Shang and Yu Zhang and Wei Liang and Anatoliy A. Savchenkov and Andrey B. Matsko and Lute Maleki and S. J. Ben Yoo},
journal = {Optica},
number = {2},
pages = {219--226},
publisher = {Optica Publishing Group},
title = {Low-loss prism-waveguide optical coupling for ultrahigh-Q low-index monolithic resonators},
volume = {5},
month = {Feb},
year = {2018},
url = {https://opg.optica.org/optica/abstract.cfm?URI=optica-5-2-219},
doi = {10.1364/OPTICA.5.000219},
}

@article{Yi:15,
author = {Xu Yi and Qi-Fan Yang and Ki Youl Yang and Myoung-Gyun Suh and Kerry Vahala},
journal = {Optica},
number = {12},
pages = {1078--1085},
publisher = {Optica Publishing Group},
title = {Soliton frequency comb at microwave rates in a high-Q silica microresonator},
volume = {2},
month = {Dec},
year = {2015},
url = {https://opg.optica.org/optica/abstract.cfm?URI=optica-2-12-1078},
doi = {10.1364/OPTICA.2.001078},
}

@Article{Herr2014,
author={Herr, T.
and Brasch, V.
and Jost, J. D.
and Wang, C. Y.
and Kondratiev, N. M.
and Gorodetsky, M. L.
and Kippenberg, T. J.},
title={Temporal solitons in optical microresonators},
journal={Nature Photonics},
year={2014},
month={Feb},
day={01},
volume={8},
number={2},
pages={145-152},
issn={1749-4893},
doi={10.1038/nphoton.2013.343},
url={https://doi.org/10.1038/nphoton.2013.343}
}

@Article{Liang2015,
author={Liang, W.
and Eliyahu, D.
and Ilchenko, V. S.
and Savchenkov, A. A.
and Matsko, A. B.
and Seidel, D.
and Maleki, L.},
title={High spectral purity Kerr frequency comb radio frequency photonic oscillator},
journal={Nature Communications},
year={2015},
month={Aug},
day={11},
volume={6},
number={1},
pages={7957},
issn={2041-1723},
doi={10.1038/ncomms8957},
url={https://doi.org/10.1038/ncomms8957}
}

@article{Xuan:16,
author = {Yi Xuan and Yang Liu and Leo T. Varghese and Andrew J. Metcalf and Xiaoxiao Xue and Pei-Hsun Wang and Kyunghun Han and Jose A. Jaramillo-Villegas and Abdullah Al Noman and Cong Wang and Sangsik Kim and Min Teng and Yun Jo Lee and Ben Niu and Li Fan and Jian Wang and Daniel E. Leaird and Andrew M. Weiner and Minghao Qi},
journal = {Optica},
number = {11},
pages = {1171--1180},
publisher = {Optica Publishing Group},
title = {High-Q silicon nitride microresonators exhibiting low-power frequency comb initiation},
volume = {3},
month = {Nov},
year = {2016},
url = {https://opg.optica.org/optica/abstract.cfm?URI=optica-3-11-1171},
doi = {10.1364/OPTICA.3.001171},
}

@article{chen_general_2012,
	title = {A general design algorithm for low optical loss adiabatic connections in waveguides},
	volume = {20},
	copyright = {© 2012 OSA},
	issn = {1094-4087},
	url = {https://opg.optica.org/oe/abstract.cfm?uri=oe-20-20-22819},
	doi = {10.1364/OE.20.022819},
	language = {EN},
	number = {20},
	urldate = {2023-05-23},
	journal = {Optics Express},
	author = {Chen, Tong and Lee, Hansuek and Li, Jiang and Vahala, Kerry J.},
	month = sep,
	year = {2012},
	note = {Publisher: Optica Publishing Group},
	pages = {22819--22829},
}

@article{yi_active_2016,
	title = {Active capture and stabilization of temporal solitons in microresonators},
	volume = {41},
	copyright = {\&\#169; 2016 Optical Society of America},
	issn = {1539-4794},
	url = {https://opg.optica.org/ol/abstract.cfm?uri=ol-41-9-2037},
	doi = {10.1364/OL.41.002037},
	language = {EN},
	number = {9},
	urldate = {2022-10-20},
	journal = {Optics Letters},
	author = {Yi, Xu and Yang, Qi-Fan and Yang, Ki Youl and Vahala, Kerry},
	month = may,
	year = {2016},
	note = {Publisher: Optica Publishing Group},
	pages = {2037--2040},
}

@article{Brasch:16,
author = {Victor Brasch and Michael Geiselmann and Martin H. P. Pfeiffer and Tobias J. Kippenberg},
journal = {Opt. Express},
number = {25},
pages = {29312--29320},
publisher = {Optica Publishing Group},
title = {Bringing short-lived dissipative Kerr soliton states in microresonators into a steady state},
volume = {24},
month = {Dec},
year = {2016},
url = {https://opg.optica.org/oe/abstract.cfm?URI=oe-24-25-29312},
doi = {10.1364/OE.24.029312},
}

@Article{pulse_herr,
author={Obrzud, Ewelina
and Lecomte, Steve
and Herr, Tobias},
title={Temporal solitons in microresonators driven by optical pulses},
journal={Nature Photonics},
year={2017},
month={Sep},
day={01},
volume={11},
number={9},
pages={600-607},
issn={1749-4893},
doi={10.1038/nphoton.2017.140},
url={https://doi.org/10.1038/nphoton.2017.140}
}

@article{aux_laser,
author = {Shuangyou Zhang and Jonathan M. Silver and Leonardo Del Bino and Francois Copie and Michael T. M. Woodley and George N. Ghalanos and Andreas {\O}. Svela and Niall Moroney and Pascal Del'Haye},
journal = {Optica},
number = {2},
pages = {206--212},
publisher = {Optica Publishing Group},
title = {Sub-milliwatt-level microresonator solitons with extended access range using an auxiliary laser},
volume = {6},
month = {Feb},
year = {2019},
url = {https://opg.optica.org/optica/abstract.cfm?URI=optica-6-2-206},
doi = {10.1364/OPTICA.6.000206},
}

@Article{aux2,
	author={Zhou, Heng
	and Geng, Yong
	and Cui, Wenwen
	and Huang, Shu-Wei
	and Zhou, Qiang
	and Qiu, Kun
	and Wei Wong, Chee},
	title={Soliton bursts and deterministic dissipative Kerr soliton generation in auxiliary-assisted microcavities},
	journal={Light: Science {\&} Applications},
	year={2019},
	month={May},
	day={29},
	volume={8},
	number={1},
	pages={50},
	issn={2047-7538},
	doi={10.1038/s41377-019-0161-y},
	url={https://doi.org/10.1038/s41377-019-0161-y}
}

@ARTICLE{fpga,
  author={Wang, Ziye and Baek, Jihoon and Ahn, Changmin and Suk, Daewon and Lee, Hansuek and Kim, Jungwon},
  journal={Journal of Lightwave Technology}, 
  title={FPGA-Driven Soliton Locking for Reduced Manual Intervention in Microcombs}, 
  year={2025},
  volume={},
  number={},
  pages={1-6},
  doi={10.1109/JLT.2025.3600657}}

@article{joshi_thermally_2016,
	title = {Thermally controlled comb generation and soliton modelocking in microresonators},
	volume = {41},
	copyright = {\&\#169; 2016 Optical Society of America},
	issn = {1539-4794},
	url = {https://opg.optica.org/ol/abstract.cfm?uri=ol-41-11-2565},
	doi = {10.1364/OL.41.002565},
	language = {EN},
	number = {11},
	urldate = {2022-09-06},
	journal = {Optics Letters},
	author = {Joshi, Chaitanya and Jang, Jae K. and Luke, Kevin and Ji, Xingchen and Miller, Steven A. and Klenner, Alexander and Okawachi, Yoshitomo and Lipson, Michal and Gaeta, Alexander L.},
	month = jun,
	year = {2016},
	note = {Publisher: Optica Publishing Group},
	pages = {2565--2568},
}

@article{rokhsari_loss_2004,
	title = {Loss characterization in microcavities using the thermal bistability effect},
	volume = {85},
	issn = {0003-6951},
	url = {https://doi.org/10.1063/1.1804240},
	doi = {10.1063/1.1804240},
	number = {15},
	urldate = {2024-11-08},
	journal = {Applied Physics Letters},
	author = {Rokhsari, H. and Spillane, S. M. and Vahala, K. J.},
	month = oct,
	year = {2004},
	pages = {3029--3031},
}

@Article{Riemensberger2020,
author={Riemensberger, Johann
and Lukashchuk, Anton
and Karpov, Maxim
and Weng, Wenle
and Lucas, Erwan
and Liu, Junqiu
and Kippenberg, Tobias J.},
title={Massively parallel coherent laser ranging using a soliton microcomb},
journal={Nature},
year={2020},
month={May},
day={01},
volume={581},
number={7807},
pages={164-170},
issn={1476-4687},
doi={10.1038/s41586-020-2239-3},
url={https://doi.org/10.1038/s41586-020-2239-3}
}

@Article{Marin-Palomo2017,
author={Marin-Palomo, Pablo
and Kemal, Juned N.
and Karpov, Maxim
and Kordts, Arne
and Pfeifle, Joerg
and Pfeiffer, Martin H. P.
and Trocha, Philipp
and Wolf, Stefan
and Brasch, Victor
and Anderson, Miles H.
and Rosenberger, Ralf
and Vijayan, Kovendhan
and Freude, Wolfgang
and Kippenberg, Tobias J.
and Koos, Christian},
title={Microresonator-based solitons for massively parallel coherent optical communications},
journal={Nature},
year={2017},
month={Jun},
day={01},
volume={546},
number={7657},
pages={274-279},
issn={1476-4687},
doi={10.1038/nature22387},
url={https://doi.org/10.1038/nature22387}
}

@Article{Feldmann2021,
author={Feldmann, J.
and Youngblood, N.
and Karpov, M.
and Gehring, H.
and Li, X.
and Stappers, M.
and Le Gallo, M.
and Fu, X.
and Lukashchuk, A.
and Raja, A. S.
and Liu, J.
and Wright, C. D.
and Sebastian, A.
and Kippenberg, T. J.
and Pernice, W. H. P.
and Bhaskaran, H.},
title={Parallel convolutional processing using an integrated photonic tensor core},
journal={Nature},
year={2021},
month={Jan},
day={01},
volume={589},
number={7840},
pages={52-58},
issn={1476-4687},
doi={10.1038/s41586-020-03070-1},
url={https://doi.org/10.1038/s41586-020-03070-1}
}

@Article{ofd,
author={Zhao, Yun
and Jang, Jae K.
and Beals, Garrett J.
and McNulty, Karl J.
and Ji, Xingchen
and Okawachi, Yoshitomo
and Lipson, Michal
and Gaeta, Alexander L.},
title={All-optical frequency division on-chip using a single laser},
journal={Nature},
year={2024},
month={Mar},
day={01},
volume={627},
number={8004},
pages={546-552},
issn={1476-4687},
doi={10.1038/s41586-024-07136-2},
url={https://doi.org/10.1038/s41586-024-07136-2}
}

@Article{Spencer2018,
author={Spencer, Daryl T.
and Drake, Tara
and Briles, Travis C.
and Stone, Jordan
and Sinclair, Laura C.
and Fredrick, Connor
and Li, Qing
and Westly, Daron
and Ilic, B. Robert
and Bluestone, Aaron
and Volet, Nicolas
and Komljenovic, Tin
and Chang, Lin
and Lee, Seung Hoon
and Oh, Dong Yoon
and Suh, Myoung-Gyun
and Yang, Ki Youl
and Pfeiffer, Martin H. P.
and Kippenberg, Tobias J.
and Norberg, Erik
and Theogarajan, Luke
and Vahala, Kerry
and Newbury, Nathan R.
and Srinivasan, Kartik
and Bowers, John E.
and Diddams, Scott A.
and Papp, Scott B.},
title={An optical-frequency synthesizer using integrated photonics},
journal={Nature},
year={2018},
month={May},
day={01},
volume={557},
number={7703},
pages={81-85},
issn={1476-4687},
doi={10.1038/s41586-018-0065-7},
url={https://doi.org/10.1038/s41586-018-0065-7}
}

@article{spectroscopy,
author = {Liron Stern  and Jordan R. Stone  and Songbai Kang  and Daniel C. Cole  and Myoung-Gyun Suh  and Connor Fredrick  and Zachary Newman  and Kerry Vahala  and John Kitching  and Scott A. Diddams  and Scott B. Papp },
title = {Direct Kerr frequency comb atomic spectroscopy and stabilization},
journal = {Science Advances},
volume = {6},
number = {9},
pages = {eaax6230},
year = {2020},
doi = {10.1126/sciadv.aax6230},
URL = {https://www.science.org/doi/abs/10.1126/sciadv.aax6230},
eprint = {https://www.science.org/doi/pdf/10.1126/sciadv.aax6230}}

@article{
spectropscopy2,
author = {Qi-Fan Yang  and Boqiang Shen  and Heming Wang  and Minh Tran  and Zhewei Zhang  and Ki Youl Yang  and Lue Wu  and Chengying Bao  and John Bowers  and Amnon Yariv  and Kerry Vahala },
title = {Vernier spectrometer using counterpropagating soliton microcombs},
journal = {Science},
volume = {363},
number = {6430},
pages = {965-968},
year = {2019},
doi = {10.1126/science.aaw2317},
URL = {https://www.science.org/doi/abs/10.1126/science.aaw2317},
eprint = {https://www.science.org/doi/pdf/10.1126/science.aaw2317}}

@Article{selfref,
author={Brasch, Victor
and Lucas, Erwan
and Jost, John D.
and Geiselmann, Michael
and Kippenberg, Tobias J.},
title={Self-referenced photonic chip soliton Kerr frequency comb},
journal={Light: Science {\&} Applications},
year={2017},
month={Jan},
day={01},
volume={6},
number={1},
pages={e16202-e16202},
issn={2047-7538},
doi={10.1038/lsa.2016.202},
url={https://doi.org/10.1038/lsa.2016.202}
}

@article{suh_searching_2019,
	title = {Searching for exoplanets using a microresonator astrocomb},
	volume = {13},
	copyright = {2018 The Author(s), under exclusive licence to Springer Nature Limited},
	issn = {1749-4893},
	url = {https://www.nature.com/articles/s41566-018-0312-3},
	doi = {10.1038/s41566-018-0312-3},
	language = {en},
	number = {1},
	urldate = {2024-11-20},
	journal = {Nature Photonics},
	author = {Suh, Myoung-Gyun and Yi, Xu and Lai, Yu-Hung and Leifer, S. and Grudinin, Ivan S. and Vasisht, G. and Martin, Emily C. and Fitzgerald, Michael P. and Doppmann, G. and Wang, J. and Mawet, D. and Papp, Scott B. and Diddams, Scott A. and Beichman, C. and Vahala, Kerry},
	month = jan,
	year = {2019},
	note = {Publisher: Nature Publishing Group},
	pages = {25--30},
}

@article{obrzud_microphotonic_2019,
	title = {A microphotonic astrocomb},
	volume = {13},
	copyright = {2018 The Author(s), under exclusive licence to Springer Nature Limited},
	issn = {1749-4893},
	url = {https://www.nature.com/articles/s41566-018-0309-y},
	doi = {10.1038/s41566-018-0309-y},
	language = {en},
	number = {1},
	urldate = {2024-11-20},
	journal = {Nature Photonics},
	author = {Obrzud, Ewelina and Rainer, Monica and Harutyunyan, Avet and Anderson, Miles H. and Liu, Junqiu and Geiselmann, Michael and Chazelas, Bruno and Kundermann, Stefan and Lecomte, Steve and Cecconi, Massimo and Ghedina, Adriano and Molinari, Emilio and Pepe, Francesco and Wildi, François and Bouchy, François and Kippenberg, Tobias J. and Herr, Tobias},
	month = jan,
	year = {2019},
	note = {Publisher: Nature Publishing Group},
	pages = {31--35},
}

@article{breath1,
  title = {Observation of Fermi-Pasta-Ulam Recurrence Induced by Breather Solitons in an Optical Microresonator},
  author = {Bao, Chengying and Jaramillo-Villegas, Jose A. and Xuan, Yi and Leaird, Daniel E. and Qi, Minghao and Weiner, Andrew M.},
  journal = {Phys. Rev. Lett.},
  volume = {117},
  issue = {16},
  pages = {163901},
  numpages = {5},
  year = {2016},
  month = {Oct},
  publisher = {American Physical Society},
  doi = {10.1103/PhysRevLett.117.163901},
  url = {https://link.aps.org/doi/10.1103/PhysRevLett.117.163901}
}

@Article{breath2,
author={Yu, Mengjie
and Jang, Jae K.
and Okawachi, Yoshitomo
and Griffith, Austin G.
and Luke, Kevin
and Miller, Steven A.
and Ji, Xingchen
and Lipson, Michal
and Gaeta, Alexander L.},
title={Breather soliton dynamics in microresonators},
journal={Nature Communications},
year={2017},
month={Feb},
day={24},
volume={8},
number={1},
pages={14569},
issn={2041-1723},
doi={10.1038/ncomms14569},
url={https://doi.org/10.1038/ncomms14569}
}

@Article{breath3,
author={Lucas, E.
and Karpov, M.
and Guo, H.
and Gorodetsky, M. L.
and Kippenberg, T. J.},
title={Breathing dissipative solitons in optical microresonators},
journal={Nature Communications},
year={2017},
month={Sep},
day={29},
volume={8},
number={1},
pages={736},
issn={2041-1723},
doi={10.1038/s41467-017-00719-w},
url={https://doi.org/10.1038/s41467-017-00719-w}
}

@article{LLE,
  title = {Spatial Dissipative Structures in Passive Optical Systems},
  author = {Lugiato, L. A. and Lefever, R.},
  journal = {Phys. Rev. Lett.},
  volume = {58},
  issue = {21},
  pages = {2209--2211},
  numpages = {0},
  year = {1987},
  month = {May},
  publisher = {American Physical Society},
  doi = {10.1103/PhysRevLett.58.2209},
  url = {https://link.aps.org/doi/10.1103/PhysRevLett.58.2209}
}

@article{Li:17,
	author = {Qing Li and Travis C. Briles and Daron A. Westly and Tara E. Drake and Jordan R. Stone and B. Robert Ilic and Scott A. Diddams and Scott B. Papp and Kartik Srinivasan},
	journal = {Optica},
	number = {2},
	pages = {193--203},
	publisher = {Optica Publishing Group},
	title = {Stably accessing octave-spanning microresonator frequency combs in the soliton regime},
	volume = {4},
	month = {Feb},
	year = {2017},
	url = {https://opg.optica.org/optica/abstract.cfm?URI=optica-4-2-193},
	doi = {10.1364/OPTICA.4.000193},
}

@article{weng_dual-mode_2022,
	title = {Dual-mode microresonators as straightforward access to octave-spanning dissipative {Kerr} solitons},
	volume = {7},
	issn = {2378-0967},
	url = {https://doi.org/10.1063/5.0089036},
	doi = {10.1063/5.0089036},
	number = {6},
	journal = {APL Photonics},
	author = {Weng, Haizhong and Afridi, Adnan Ali and Li, Jing and McDermott, Michael and Tu, Huilan and Barry, Liam P. and Lu, Qiaoyin and Guo, Weihua and Donegan, John F.},
	month = jun,
	year = {2022},
	note = {\_eprint: https://pubs.aip.org/aip/app/article-pdf/doi/10.1063/5.0089036/16492580/066103\_1\_online.pdf},
	pages = {066103},
}

@Article{Shen2020,
	author={Shen, Boqiang
	and Chang, Lin
	and Liu, Junqiu
	and Wang, Heming
	and Yang, Qi-Fan
	and Xiang, Chao
	and Wang, Rui Ning
	and He, Jijun
	and Liu, Tianyi
	and Xie, Weiqiang
	and Guo, Joel
	and Kinghorn, David
	and Wu, Lue
	and Ji, Qing-Xin
	and Kippenberg, Tobias J.
	and Vahala, Kerry
	and Bowers, John E.},
	title={Integrated turnkey soliton microcombs},
	journal={Nature},
	year={2020},
	month={Jun},
	day={01},
	volume={582},
	number={7812},
	pages={365-369},
	issn={1476-4687},
	doi={10.1038/s41586-020-2358-x},
	url={https://doi.org/10.1038/s41586-020-2358-x}
}

@article{Wildi:19,
	author = {Thibault Wildi and Victor Brasch and Junqiu Liu and Tobias J. Kippenberg and Tobias Herr},
	journal = {Opt. Lett.},
	number = {18},
	pages = {4447--4450},
	publisher = {Optica Publishing Group},
	title = {Thermally stable access to microresonator solitons via slow pump modulation},
	volume = {44},
	month = {Sep},
	year = {2019},
	url = {https://opg.optica.org/ol/abstract.cfm?URI=ol-44-18-4447},
	doi = {10.1364/OL.44.004447},
}

@article{ssb,
	title = {Thermal and Nonlinear Dissipative-Soliton Dynamics in Kerr-Microresonator Frequency Combs},
	author = {Stone, Jordan R. and Briles, Travis C. and Drake, Tara E. and Spencer, Daryl T. and Carlson, David R. and Diddams, Scott A. and Papp, Scott B.},
	journal = {Phys. Rev. Lett.},
	volume = {121},
	issue = {6},
	pages = {063902},
	numpages = {6},
	year = {2018},
	month = {Aug},
	publisher = {American Physical Society},
	doi = {10.1103/PhysRevLett.121.063902},
	url = {https://link.aps.org/doi/10.1103/PhysRevLett.121.063902}
}

@Article{Kudelin2024,
	author={Kudelin, Igor
	and Groman, William
	and Ji, Qing-Xin
	and Guo, Joel
	and Kelleher, Megan L.
	and Lee, Dahyeon
	and Nakamura, Takuma
	and McLemore, Charles A.
	and Shirmohammadi, Pedram
	and Hanifi, Samin
	and Cheng, Haotian
	and Jin, Naijun
	and Wu, Lue
	and Halladay, Samuel
	and Luo, Yizhi
	and Dai, Zhaowei
	and Jin, Warren
	and Bai, Junwu
	and Liu, Yifan
	and Zhang, Wei
	and Xiang, Chao
	and Chang, Lin
	and Iltchenko, Vladimir
	and Miller, Owen
	and Matsko, Andrey
	and Bowers, Steven M.
	and Rakich, Peter T.
	and Campbell, Joe C.
	and Bowers, John E.
	and Vahala, Kerry J.
	and Quinlan, Franklyn
	and Diddams, Scott A.},
	title={Photonic chip-based low-noise microwave oscillator},
	journal={Nature},
	year={2024},
	month={Mar},
	day={01},
	volume={627},
	number={8004},
	pages={534-539},
	issn={1476-4687},
	doi={10.1038/s41586-024-07058-z},
	url={https://doi.org/10.1038/s41586-024-07058-z}
}

@Article{Sun2024,
	author={Sun, Shuman
	and Wang, Beichen
	and Liu, Kaikai
	and Harrington, Mark W.
	and Tabatabaei, Fatemehsadat
	and Liu, Ruxuan
	and Wang, Jiawei
	and Hanifi, Samin
	and Morgan, Jesse S.
	and Jahanbozorgi, Mandana
	and Yang, Zijiao
	and Bowers, Steven M.
	and Morton, Paul A.
	and Nelson, Karl D.
	and Beling, Andreas
	and Blumenthal, Daniel J.
	and Yi, Xu},
	title={Integrated optical frequency division for microwave and mmWave generation},
	journal={Nature},
	year={2024},
	month={Mar},
	day={01},
	volume={627},
	number={8004},
	pages={540-545},
	issn={1476-4687},
	doi={10.1038/s41586-024-07057-0},
	url={https://doi.org/10.1038/s41586-024-07057-0}
}

@article{Qi:19,
	author = {Zhen Qi and Shaokang Wang and Jos\'{e} Jaramillo-Villegas and Minghao Qi and Andrew M. Weiner and Giuseppe D'Aguanno and Thomas F. Carruthers and Curtis R. Menyuk},
	journal = {Optica},
	number = {9},
	pages = {1220--1232},
	publisher = {Optica Publishing Group},
	title = {Dissipative cnoidal waves (Turing rolls) and the soliton limit in microring resonators},
	volume = {6},
	month = {Sep},
	year = {2019},
	url = {https://opg.optica.org/optica/abstract.cfm?URI=optica-6-9-1220},
	doi = {10.1364/OPTICA.6.001220},
}

\end{document}